\documentclass[a4paper,11pt]{article}

 \usepackage[latin1]{inputenc}
 \usepackage[T1]{fontenc}
 \usepackage[american]{babel}
 \usepackage[dvips]{graphics,graphicx,color}
 \usepackage[bf]{subfigure}
 \usepackage[hang,footnotesize,bf]{caption2} 
\begin{document}

\begin{center}
  
  \hbox{}\vspace{0.5cm}
  {\Large \bf Analytical procedure to determine the self-referred lacunarity
  function for simple shapes. Supplementary material.}\\
  \vspace{0.5cm} {\sf } 
  {\sf Erbe P. Rodrigues, Marconi. S. Barbosa and Luciano da F. Costa}\\
  {\small Instituto de Física de São Carlos \\
  Universidade de São Paulo Caixa Postal 369 CEP 13560.970 São Carlos, SP, Brazil}\\
  {\sl e-mail: marconi@if.sc.usp.br}\\

\end{center}

\vspace{.5cm}

\begin{abstract}
  The analytical calculation of the self-referred lacunarity is used as a
  validation standard of the computational algorithm. In this supplementary
  material to our article (see cond-mat/0407079) we present a detailed
  calculation for two simple shapes, namely a square box and a cross.
\end{abstract}

\section{Introduction: self-referred lacunarity}
The lacunarity of an object is defined as 
\[\Lambda(r)=\frac{Z^{(2)}}{[Z{(1)}]^2}=\frac{\sigma^2(r)}{[\mu(r)]^2}+1,\]
where $Z^{(1)}$ and $Z^{(2)}$ are the first and second moments of a mass
distribution and $\sigma^2(r)$ and $\mu(r)$ are its variance and mean.  We can
calculate $\mu(r)$ and $\sigma^2(r)$ by defining a function $A(x,y,r)$ that gives
the area of the object inside a window of radius $r$, whose center is in the
point $(x,y)$, in the following way
\begin{eqnarray} 
\mu(r)&=&\frac{\int_x\int_yA(x,y,r)dxdy}{\int_x\int_ydxdy}  \label{eq_mean}, \\
\sigma^2(r)&=&\frac{\int_x\int_y[\mu(r)-A(x,y,r)]^2dxdy}{\int_x\int_ydxdy}. \label{eq_vari} 
\end{eqnarray} 
For the self-referred approach of lacunarity the integral is calculated only
for values of $x$ and $y$ that fall over the object, see Figure \ref{l_a_quad}.
 
\section{Analytical calculation for simple shapes}
\subsection{Object: a square box} 
\begin{figure}[htb]
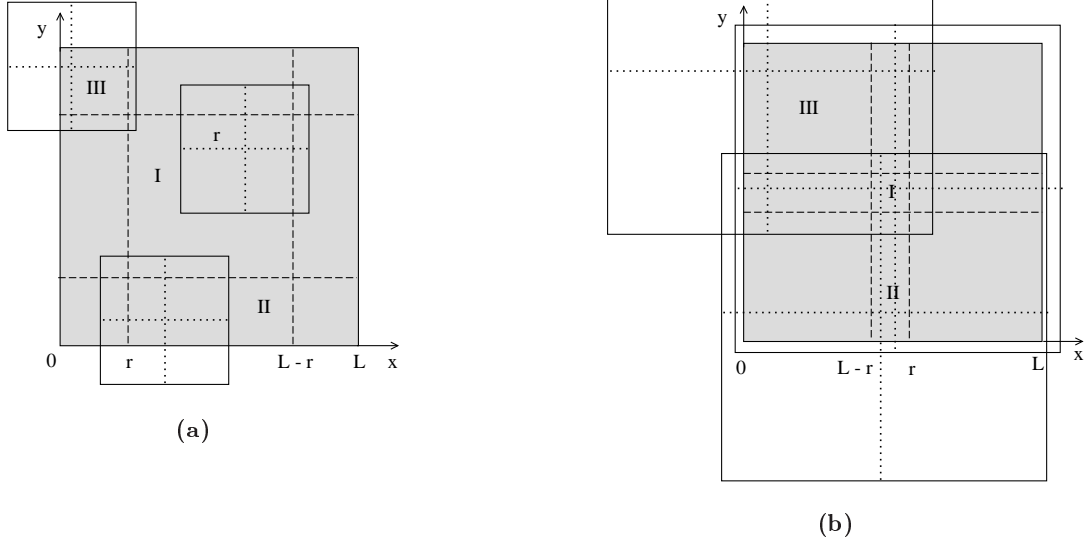

\centering
   \subfigure[]{
      \label{l_a_quad_a}
      \begin{minipage}[c]{0.45\linewidth}
         \centering
         \includegraphics[scale=.45,clip]{lac_analitica_quadrado.eps}
      \end{minipage}}
        \hfill  
        \subfigure[]{
      \label{l_a_quad_b}
      \begin{minipage}[c]{0.5\linewidth}
         \centering
         \includegraphics[scale=.45,clip]{lac_analitica_quadrado_II.eps}
      \end{minipage}}
\caption{ Square regions that correspond to the self-referred lacunarity
  calculation for the intervals (a) $r\in[0,L/2]$ and (b) $r\in[L/2,L]$. The
  sliding windows in the region $I$ are completely filled in (a) and capture
  all the square in (b). In the regions $II$ and $III$ they are partially
  filled in (a) as well as in (b).}
\label{l_a_quad}
\end{figure}

We first show how to determine the self-referred lacunarity for a solid square
of side $L$ shown in Figure~\ref{l_a_quad_a}. In order to calculate the
analytical lacunarity of a square it is necessary to determine the area
funcition $A(x,y,r)$.  This function is determined by dividing the square in
regions labeled by $I$, $II$ and $III$ as in Figure~\ref{l_a_quad_a}
and~\ref{l_a_quad_b}.  There is one region of type $I$ and four of type $II$
and $III$. It is also necessary to divide this calculation into two intervals:
$r=[0,L/2]$ and $r=[L/2,L]$.

\begin{figure}[!htb]
\centering
\includegraphics[scale=.5,clip]{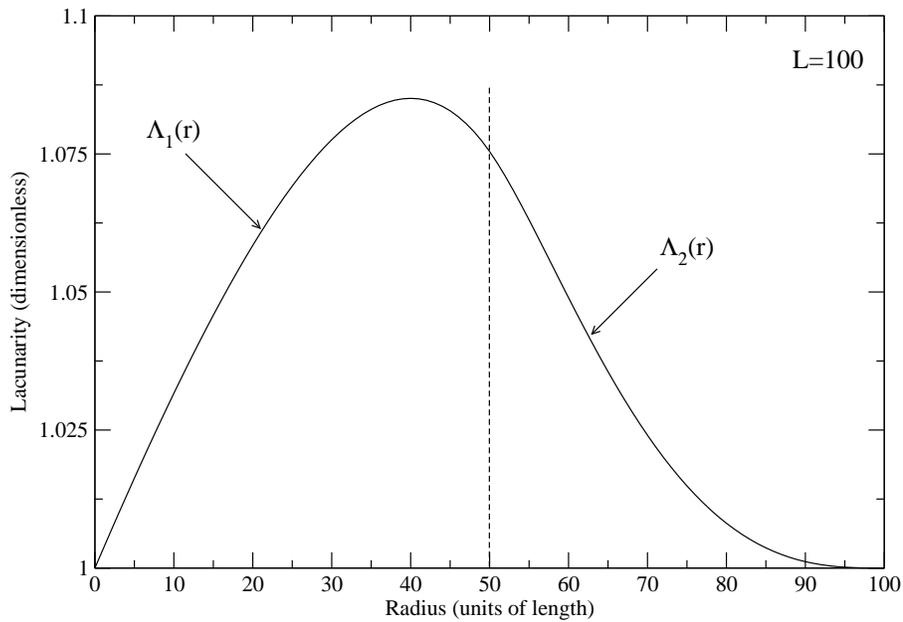} 
\caption{The analytically calculated self-referred lacunarity curve for a
  square with side $L$ of $100$ absolute units of length.}  
\label{inv_esc_quad}
\end{figure}

For the first interval, $r\in[0,L/2]$, the functions $A^1_i(x,y,r)$ and
respective integration ranges for $x$ and $y$ are
\begin{eqnarray} \nonumber
A^1_{I}(x,y,r)&=&4r^2,\quad x\in[r,(L-r)],\quad \in[r,(L-r)] \\ \nonumber
A^1_{II}(x,y,r)&=&2r^2+2ry,\quad x\in[r,(L-r)],\quad \in[0,r] \\ \nonumber
A^1_{III}(x,y,r)&=&r^2+[r+(L-y)]x+r(L-y),  \\ \nonumber
                                && x\in[0,r],\quad y\in[(L-r),r] \nonumber
\end{eqnarray} 
The mean value is determined by the integration of the $A^1_i$ over their
respective regions divided by the total area of the square, according the
Equation~\ref{eq_mean}
\begin{eqnarray}
\mu_1(r)&=&\frac{\int_x\int_y\sum_iA^1_i(x,y,r)dxdy}{\int_x\int_ydxdy} \nonumber \\
                &=&\frac{1}{L^2}\int\int A^1_{I}(x,y,r)+4A^1_{II}(x,y,r)+4A^1_{III}(x,y,r)dxdy  
\nonumber \\
                &=&\frac{r^2(r-2L)^2}{L^2} \nonumber 
\end{eqnarray} 
The four regions $II$ and $III$ are equivalent and it is necessary only to
integrate over one of region $II$ and $III$ and multiply the result by 4.  The
variance is calculated in analogous way according to Equation~\ref{eq_vari}
\begin{eqnarray} 
        \sigma_1^2(r) &=& \frac{1}{L^2} \int\int (A^1_{I}(x,y,r)-\mu_1(r))^2 \\ \nonumber
                            &+&   4(A^1_{II}(x,y,r)-\mu_1(r))^2  \\ \nonumber
                            &+&   4(A^1_{III}(x,y,r)-\mu_1(r))^2 dxdy \\ \nonumber  
                                &=&     -\frac{r^5(r-6L)(3r-4L)(3r-2L)}{9L^4} 
\end{eqnarray} 
Knowing $\mu_1(r)$ and $\sigma_1^2(r)$ we may determine the expression for
self-referred lacunarity for the first interval
\[\Lambda_1(r)=\frac{\sigma_1^2(r)}{[\mu_1(r)]^2}+1=\frac{4L^2(5r-6L)^2}{9(r-2L)^4}\]
The calculation for the second interval, $r\in[L/2,L]$, proceeds in the same way,
with the $A^2_i$ functions and their respectives integration intervals given by
\begin{eqnarray} 
A^2_{I}(x,y,r)&=& L^2,\quad x\in[(L-r),r],\quad y\in[(L-r),r]  \nonumber \\
A^2_{II}(x,y,r)&=& L(r+y),\quad x\in[(L-r),r],\quad y\in[0,(L-r)] \nonumber \\
A^2_{III}(x,y,r)&=& r^2+[r+(L-y)]x+r(L-y),  \nonumber \\
                                && x\in[0,(L-r)],\quad y\in[r,L] \nonumber
\end{eqnarray} 
and the mean, variance and Lacunarity as in  
\begin{eqnarray} 
\mu_2(r)&=&\frac{r^2(r-2L)^2}{L^2}  \nonumber \\
\sigma_2^2(r)&=&-\frac{(3L-r)(r-L)^3(3r^4-14r^3L+12r^2L^2+6rL^3-L^4)}{9L^4} \nonumber \\ 
\Lambda_2(r) &=&\frac{L^2(2r^3-6rL^2+L^3)^2}{9r^4(r-2L)^4} \nonumber
\end{eqnarray} 
The final expression for the whole interval self-referred lacunarity is
\begin{eqnarray} 
 \Lambda(r) = \left\{ \begin{array}{ll} \frac{4L^2(5r-6L)^2}{9(r-2L)^4}, &  r \in [0,L/2] ; \\ \\ \frac{L^2(2r^3-6rL^2+L^3)^2}{9r^4(r-2L)^4}, &  r \in [L/2,L]. \end{array} \right. \nonumber
\end{eqnarray}
Figure~\ref{inv_esc_quad} shows a plot of the expression above for a
square of length $L=100$, in absolute units of length.

\subsection{Object: a cross} 
\begin{figure}[!ht]
\centering
\includegraphics[scale=.55,clip]{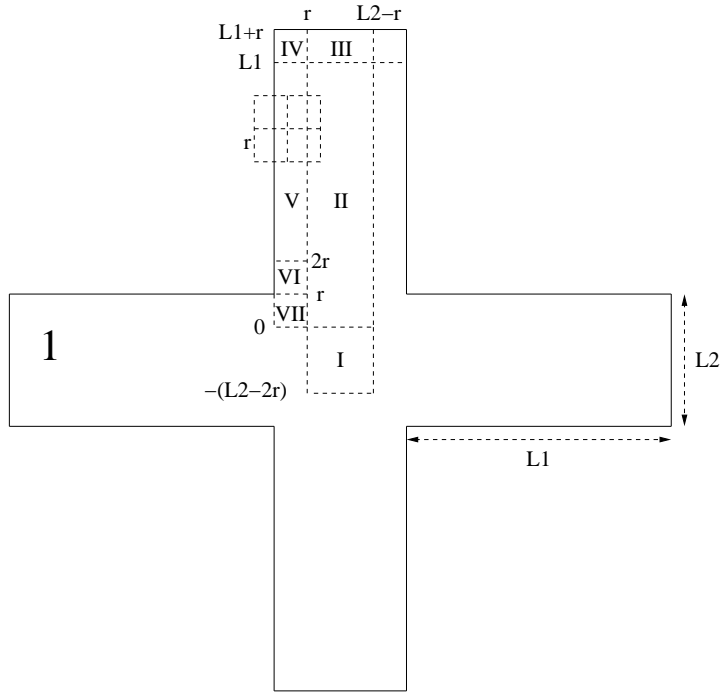} 
\caption{The cross divided to regions for the first interval.}
\label{cruz1}
\end{figure}

\begin{figure}[!htb]
\centering
\includegraphics[scale=.55,clip]{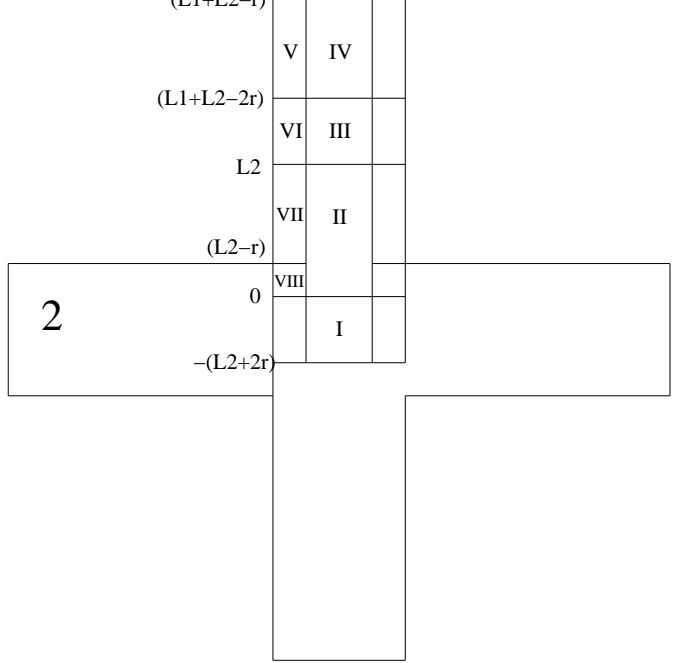}
\caption{The cross regions to interval $\frac{L2}{2} \leq r\leq L2$.}
\label{cruz2}
\end{figure}

\begin{figure}[!htb]
\centering
\includegraphics[scale=.45,clip]{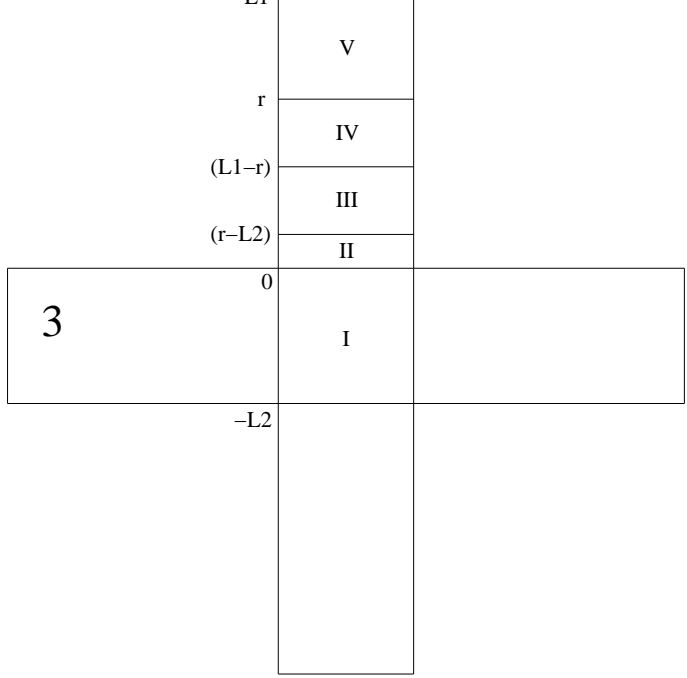}
\caption{The cross regions for the interval $L2\leq r\leq\frac{(L1+L2)}{2}$. }
\label{cruz3}
\end{figure}

\begin{figure}[!htb]
\centering
\includegraphics[scale=.45,clip]{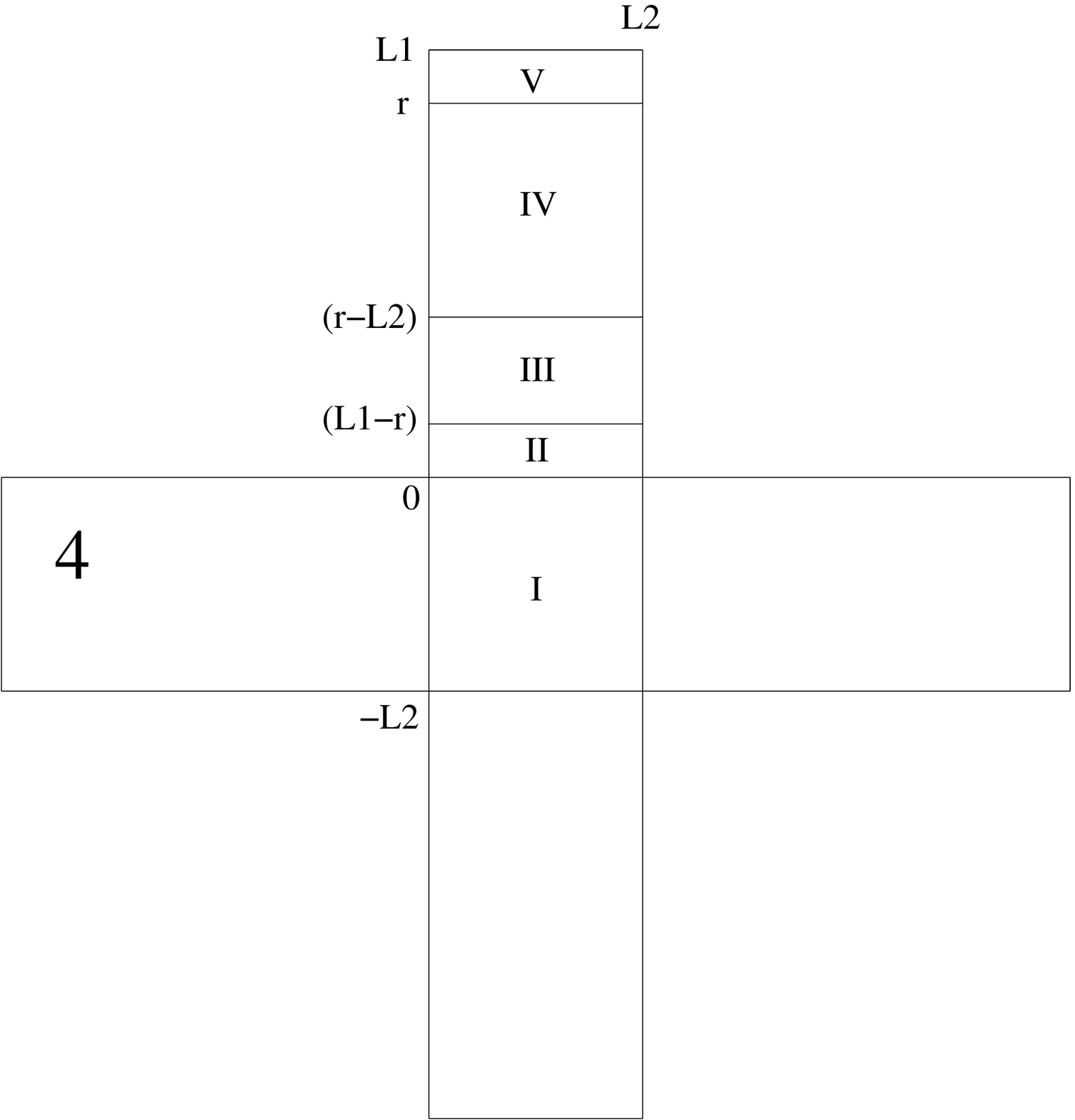}
\caption{The cross regions for the interval $\frac{(L1+L2)}{2}\leq r \leq L1$. }
\label{cruz4}
\end{figure}

\begin{figure}[!htb]
\centering
\includegraphics[scale=.45,clip]{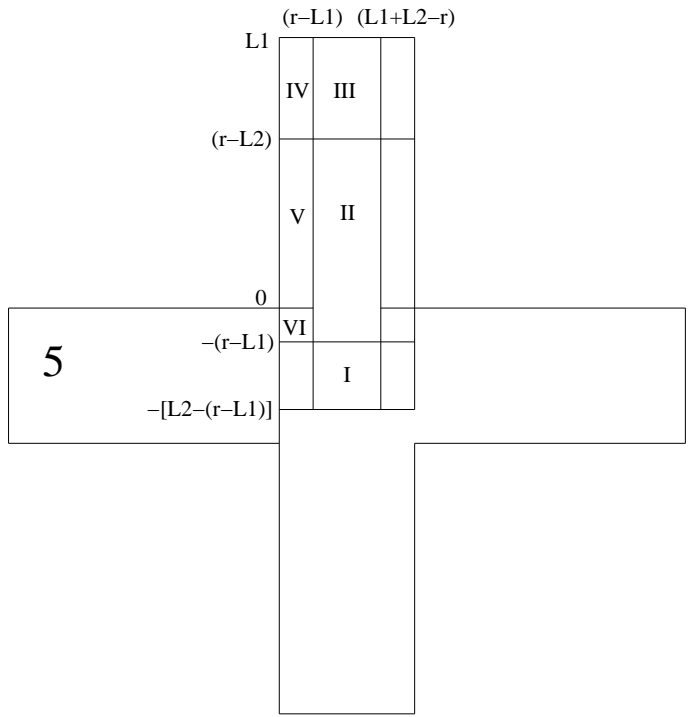}
\caption{The cross regions for the interval $L1\leq r\leq L1+\frac{L2}{2}$. }
\label{cruz5}
\end{figure}

\begin{figure}[!htb]
\centering
\includegraphics[scale=.45,clip]{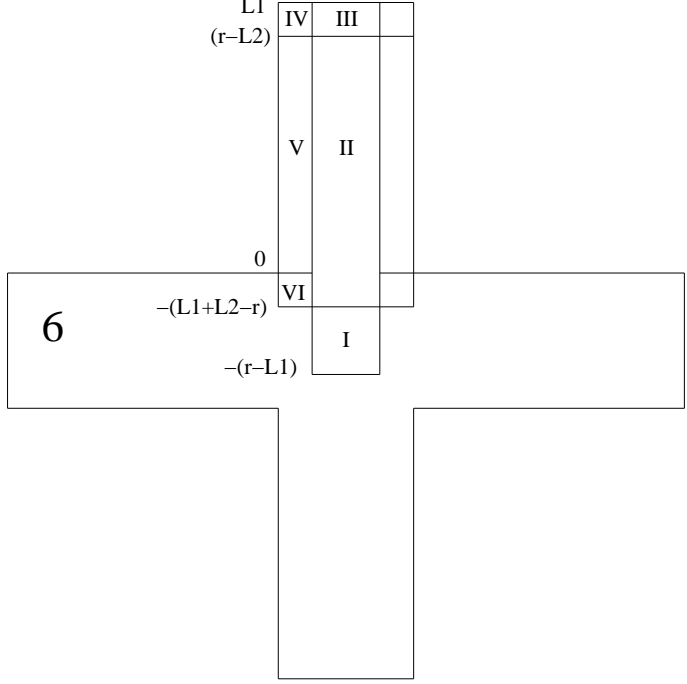}
\caption{The cross regions for the interval $L1+\frac{L2}{2}\leq r \leq L1+L2$. }
\label{cruz6}
\end{figure}

\begin{figure}[!htb]
\centering
\includegraphics[scale=.45,clip]{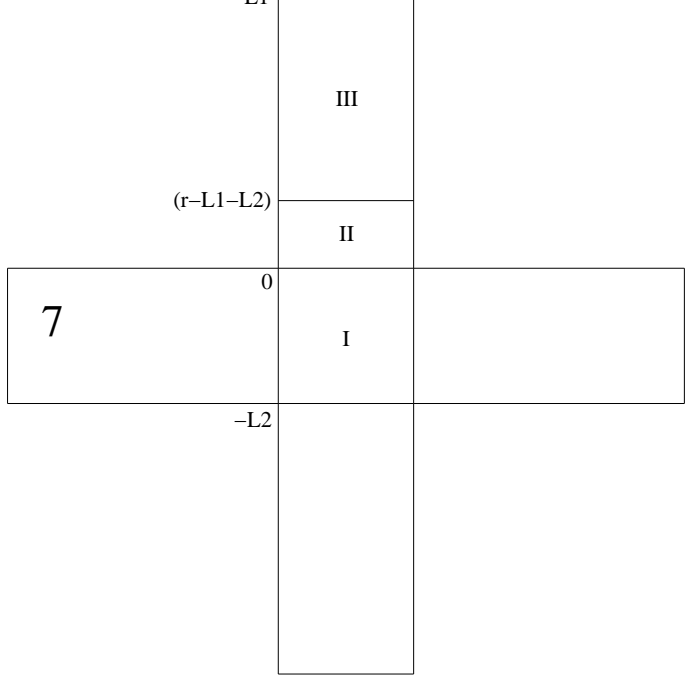}
\caption{The cross regions for the interval $L1+L2\leq r\leq 2L1+L2$.}
\label{cruz7}
\end{figure}

The analytical procedure to calculate the lacunarity of a cross, which
occupies a region $k$, is similar to the procedure presented above for a
square. Again we define a function $A(x,y,r)$.  For this case, there are seven
intervals and each one was divided into regions which define an associated
function $A_i(x,y,r)$. Figure~\ref{cruz1} shows the first interval considered
in the self-referred lacunarity calculation of a cross with dimensions $L1$
and $L2$ and the respective regions. Figure~\ref{cruz1} presents only
non-repeated regions $i$, however the calculation regards all the possible
regions $j$. In the following we show the calculation of the first interval in
more detail. For the remaining intervals we provide only the functions
$A_i(x,y,r)$, $\mu(r)$, $\sigma^2(r)$ and $\Lambda(r)$. Figures~\ref{cruz2}
to~\ref{cruz7} presents the divided regions for the remaining intervals
labeled from $2$ to $7$. The expression for lacunarity,
$\Lambda(r)=(\sigma^2(r)/ \mu^2(r))+1$ where $\mu$ and $\sigma^2$ is given by
\[\mu(r)=\frac{\int_k\sum A_j(x,y,r)dk}{\int_kdk},\] 
and
\[\sigma^2(r)=\frac{\int_k\left[\sum A_j(x,y,r)-\mu(r)\right]^2dk}{\int_kdk},\] 
where $\int_kdk$ is the total $k$ area ($A_k=4L1L2+L2^2$) of the cross.

\subsubsection{Interval $0 \leq r \leq \frac{L2}{2}$}
\begin{eqnarray}
A_{I}(x,y,r)  &=& 4r^2  \quad x\in[r,L2-r], \quad y\in[-(L2-2r),0] \nonumber \\
A_{II}(x,y,r) &=& 4r^2  \quad x\in[r,L2-r], \quad y\in[0,L1] \nonumber \\
A_{III}(x,y,r)&=& (2r+L1-y)2r    \quad x\in[r,L2-r], \quad y\in[L1,L1+r] \nonumber \\
A_{IV}(x,y,r) &=& (r+x)(2r+L1-y) \quad x\in[0,r], \quad y\in[L1,L1+r] \nonumber \\
A_{V}(x,y,r)  &=& (r+x)(2r)      \quad x\in[0,r], \quad y\in[2r,L1] \nonumber \\
A_{VI}(x,y,r) &=& 4r^2-y(r-x)    \quad x\in[0,r], \quad y\in[r,2r] \nonumber \\
A_{VII}(x,y,r)&=& 4r^2-y(r-x)    \quad x\in[0,r], \quad y\in[0,r] \nonumber 
\end{eqnarray}
The Calculation of the mean $\mu$ for the first region gives
\begin{eqnarray}
\mu_I(r)&=&\frac{\int_k(A_I+4A_{II}+4A_{III}+8A_{IV}+8A_{V}+8A_{VI}+4A_{VII})dk}{A_k}\nonumber \\
                &=&\frac{r^2(4L2^2-4rL2+3r^2+16L1L2-8rL1)}{L2(4L1+L2)}. \nonumber
\end{eqnarray}
We make $L1=2L$ e $L2=L$ to simplify the expressions. Thus we have for the previous expression
\[\mu_I(r)=\frac{r^2(36l^2-20Lr+3r^2)}{9L^2}.\]
In an analogous way $\sigma^2(r)$ is given by
\begin{eqnarray}
\sigma^2_I(r)&=&\frac{\int_k\left(\left[A_I-\mu_I(L1,L2,r)\right]^2+\ldots+
                                                   \left[4A_{VII}-\mu_I(L1,L2,r)\right]^2\right)dk}{A_k}
\nonumber \\
&=&-\frac{1}{9L2^2(4L1+L2)^2}(-384L2L1^2+576rL1^2-432r^2L1-288L1L2^2 \nonumber \\
& & + 496rL1L2-216L2r^2-48L2^3+81r^3+124rL2^2)r^5 .\nonumber
\end{eqnarray} 
Simplifying the above expression we have
\[\sigma^2_I(r)=-\frac{{r}^{5} \left( 9{r}^{3}-120L{r}^{2}+380
r{L}^{2}-240{L}^{3} \right) }{81L^{4}}.\]
The lacunarity expression in this interval is finally given by
\[\Lambda_I(r)=\frac{4L^2(-300Lr+59r^2+324L^2)}{(36L^{2}-20Lr+3r^2)^2}.\]
\\
\\
\subsubsection{Interval $\frac{L2}{2} \leq r \leq L2$}
\begin{eqnarray}
A_{I}(x,y,r)  &=&  2(2rL2)-L2^2  \quad x\in[L2-r,r],y\in[-(L2+2r),0] \nonumber \\
A_{II}(x,y,r) &=& (2r+y-L2)L2+(L2-y)2r\quad x\in[L2-r,r],\quad y\in[0,L2] \nonumber \\
A_{III}(x,y,r)&=& L2 \times 2r \quad x\in[L2-r,r], \quad y\in[L2,L1+L2-2r] \nonumber \\
A_{IV}(x,y,r) &=& L2(L1+L2-y)    \quad x\in[L2-r,r],\nonumber \\
&&y\in[L1+L2-2r,L1+L2-r] \nonumber \\
A_{V}(x,y,r)  &=&  (L1+L2-y)(r+x) \quad x\in[0,L2-r],\nonumber \\
&& y\in[L1+L2-2r,L1+L2-r] \nonumber \\
A_{VI}(x,y,r) &=& 2r(r+x)    \quad x\in[0,L2-r], \quad y\in[L2,L1+L2-2r] \nonumber \\
A_{VII}(x,y,r)&=& 2r(r+x)+(r-x)(L2-y)    \quad x\in[0,L2-r],y\in[L2-r,L2]\nonumber \\
A_{VIII}(x,y,r)&=& 4r^2-(2r+y-L2)(r-x) \quad x\in[0,L2-r], \quad y\in[0,L2-r] \nonumber 
\end{eqnarray}

\[\mu_{II}(r)=\frac{r^2(-20rL+36L^2+3r^2)}{9L^2}\]
\begin{eqnarray}
\sigma^2_{II}(r)&=&\frac{1}{81L^4}(120r^7L-12L^7r-9r^8+1248r^5L^3+452L^5r^3 \nonumber \\ 
&&+L^8-36L^6r^2-604r^6L^2-1128L^4r^4)\nonumber
\end{eqnarray}

\[\Lambda_{II}(r)={\frac {{L}^{2} \left( -12\,{L}^{5}r-192\,{r}^{5}L+452\,{L}^{3}{r}^{3}
+{L}^{6}-36\,{L}^{4}{r}^{2}+12\,{r}^{6}+168\,{r}^{4}{L}^{2} \right) }{
{r}^{4} \left( -20\,rL+36\,{L}^{2}+3\,{r}^{2} \right) ^{2}}}\]
\\
\\
\subsubsection{Interval $L2 \leq r \leq \frac{L1+L2}{2}$}
\begin{eqnarray}
A_{I}(x,y,r)  &=&  2L2(2r)-L2^2  \quad x\in[0,L2],\quad y\in[-L2,0] \nonumber \\
A_{II}(x,y,r) &=& 4rL2-L2^2 \quad x\in[0,L2],\quad y\in[0,r-L2] \nonumber \\
A_{III}(x,y,r)&=& L2(r+y)+(r-y)2r \quad x\in[0,L2], \quad y\in[r-L2,L1-r] \nonumber \\
A_{IV}(x,y,r) &=& (L1L2)+(r-y)2r \quad x\in[0,L2], \quad y\in[L1-r,r)] \nonumber \\
A_{V}(x,y,r)  &=& (L1-(y-r))L2 \quad x\in[0,L2], \quad y\in[r,L1] \nonumber 
\end{eqnarray}

\[\mu_{III}(r)=\frac{4L}{3}rL+\frac{1}{9}L^2+\frac{2}{3}{r}^{2}\]

\[\sigma^2_{III}(r)=-{\frac {76}{27}}\,{r}^{4}-{\frac {124}{9}}\,{r}^{2}{L}^{2}+{\frac {
100}{9}}\,L{r}^{3}+{\frac {196}{27}}\,{L}^{3}r-{\frac {112}{81}}\,{L}^
{4}\]

\[\Lambda_{III}(r)=-3\,{\frac {64\,{r}^{4}+320\,{r}^{2}{L}^{2}-348\,L{r}^{3}-204\,{L}^{3}
r+37\,{L}^{4}}{ \left( 12\,rL+{L}^{2}+6\,{r}^{2} \right) ^{2}}}\]
\\
\\
\subsubsection{Interval $\frac{L1+L2}{2} \leq r \leq L1$}
\begin{eqnarray}
A_{I}(x,y,r)  &=& 2L2(2r)-L2^2\quad x\in[0,L2],\quad y\in[-L2,0] \nonumber \\
A_{II}(x,y,r) &=& 2rL2+(r-y-L2)L2+(y+r)L2\quad x\in[0,L2],\quad y\in[0,L1-r] \nonumber \\
A_{III}(x,y,r)&=& L1L2+2rL2+(r-y-L2)L2\quad x\in[0,L2], \quad y\in[L1-r,r-L2] \nonumber \\
A_{IV}(x,y,r) &=& L1L2+(r-y)2r\quad x\in[0,L2], \quad y\in[r-L2,r)] \nonumber \\
A_{V}(x,y,r)  &=& (L1-(y-r))L2\quad x\in[0,L2], \quad y\in[r,L1] \nonumber 
\end{eqnarray}

\[\mu_{IV}(r)=\frac{4L}{3}rL+\frac{1}{9}L^2+\frac{2}{3}{r}^{2}\]

\[\sigma^2_{IV}(r)=-\frac{4}{9}\,{r}^{4}+{\frac {68}{9}}\,{r}^{2}{L}^{2}-{\frac {20}{27}}\,L{r}^{
3}-{\frac {236}{27}}\,{L}^{3}r+{\frac {212}{81}}\,{L}^{4}\]

\[\Lambda_{IV}(r)=3\,{\frac {L \left( 256\,{r}^{2}L+28\,{r}^{3}-228\,r{L}^{2}+71\,{L}^{3
} \right) }{ \left( 12\,rL+{L}^{2}+6\,{r}^{2} \right) ^{2}}}\]
\\
\\
\subsubsection{Interval $L1 \leq r \leq L1+ \frac{L2}{2}$}
\begin{eqnarray}
A_{I}(x,y,r)  &=& 2(2rL2)-L2^2\quad x\in[r-L1,L1+L2-r],\nonumber \\
&& y\in[-L1-L2+r,-r+L1] \nonumber \\
A_{II}(x,y,r) &=&L1L2+2rL2+(r-y-L2)L2\quad x\in[r-L1,L2+L1-r], \nonumber \\
&& y\in[-r+L1,r-L2] \nonumber \\
A_{III}(x,y,r)&=& L1L2)+(r-y)2r\quad x\in[r-L1,L2+L1-r],\quad y\in[r-L2,L1] \nonumber \\
A_{IV}(x,y,r) &=& L1L2+(r-y)(r+x+L1)\quad x\in[0,r-L1],\quad y\in[r-L2,L1] \nonumber \\
A_{V}(x,y,r)  &=& L1L2+(r+x+L1)L2+(r-y-L2)L2\quad x\in[0,r-L1],\nonumber \\
&& y\in[0,r-L2] \nonumber \\
A_{VI}(x,y,r) &=& L1L2+(r+x+L1)L2+(r-y-L2)L2\quad x\in[0,r-L1],\nonumber \\ 
&& y\in[-(r-L1),0] \nonumber 
\end{eqnarray}

\[\mu_{V}(r)=\frac {-100\,{L}^{3}r+90\,{r}^{2}{L}^{2}-24\,L{r}^{3}+2\,{r}^{4}
+49\,{L}^{4}}{9{L}^{2}}\]

\begin{eqnarray}
\sigma^2_{V}(r)&=&-\frac{2}{81L^4}(398\,{L}^{8}+6704\,{L}^{6}{r}^{2}+5717\,{L}^
{4}{r}^{4}+448\,{r}^{6}{L}^{2} \nonumber \\
               & & -2144\,{r}^{5}{L}^{3}-8516\,{L}^{5}{r}^{3}-2542\,{L}^{7}r-48\,{r}^{7}L+2\,{r}^{8}) \nonumber
\end{eqnarray}

\[\Lambda_{V}(r)=\frac {{L}^{2} \left( 1605\,{L}^{6}+5412\,{L}^{4}{r}^{2}+1662\,{r}^{4
}{L}^{2}+40\,{r}^{6}-432\,{r}^{5}L-3320\,{L}^{3}{r}^{3}-4716\,{L}^{5}r
 \right) }{ \left( -100\,{L}^{3}r+90\,{r}^{2}{L}^{2}-24\,L{r}^{3}+2\,{
r}^{4}+49\,{L}^{4} \right) ^{2}}\]
\\
\\
\subsubsection{Interval $ L1 + \frac{L2}{2} \leq r \leq L1 + L2$}
\begin{eqnarray}
A_{I}(x,y,r)  &=& 4L1L2+L2^2\quad x\in[L1+L2-r,r-L1],\nonumber \\
&& y\in[-(r-L1),-(L1+L2-r)] \nonumber \\
A_{II}(x,y,r) &=& L2^2+3L1L2+(r-y-L2)L2\quad x\in[-(L2+L1-r),r-L1], \nonumber \\
&& y\in[-(L1+L2-r),r-L2] \nonumber \\
A_{III}(x,y,r)&=& L1L2+(r-y)(2L1+L2)\quad x\in[L2+L1-r,r-L1],\nonumber \\
&& y\in[r-L2,L1] \nonumber \\
A_{IV}(x,y,r) &=& L1L2+(L1+r+x)(r-y)\quad x\in[0,L2+L1-r],\quad y\in[r-L2,L1] \nonumber \\
A_{V}(x,y,r)  &=& L1L2+(r-y-L2)L2+(L1+r+x)L2\quad x\in[0,L2+L1-r],\nonumber \\
&& y\in[0,r-L2] \nonumber \\
A_{VI}(x,y,r) &=& L1L2+(r-y-L2)L2+(L1+r+x)L2\quad x\in[0,L2+L1-r],\nonumber \\
&& y\in[-(L1+L2-r),0] \nonumber 
\end{eqnarray}

\[\mu_{VI}(r)=\frac {-100\,{L}^{3}r+90\,{r}^{2}{L}^{2}-24\,L{r}^{3}+2\,{r}^{4}
+49\,{L}^{4}}{9{L}^{2}}\]

\begin{eqnarray}
\sigma^2_{VI}(r)&=&-\frac {2}{81L^4}(11804\,{L}^{6}{r}^{2}+6881\,{L}^{4}{r}^{4}+
464\,{r}^{6}{L}^{2}-2360\,{r}^{5}{L}^{3} \nonumber \\ 
& &-11766\,{L}^{5}{r}^{3}-7042\,{L}^{7}r-48\,{r}^{7}L+2\,{r}^{8}+2273\,{L}^{8})\nonumber
\end{eqnarray}

\[\Lambda_{VI}(r)=-{\frac {{L}^{2} \left( 4788\,{L}^{4}{r}^{2}+666\,{r}^{4}{L}^{2}-8\,{r
}^{6}-3180\,{L}^{3}{r}^{3}-4284\,{L}^{5}r+2145\,{L}^{6} \right) }{
 \left( -100\,{L}^{3}r+90\,{r}^{2}{L}^{2}-24\,L{r}^{3}+2\,{r}^{4}+49\,
{L}^{4} \right) ^{2}}}\]
\\
\\
\subsubsection{Interval $ L1+L2 \leq r \leq 2L1+L2$}
\begin{figure}[!htb]
\centering
\includegraphics[scale=.5,clip]{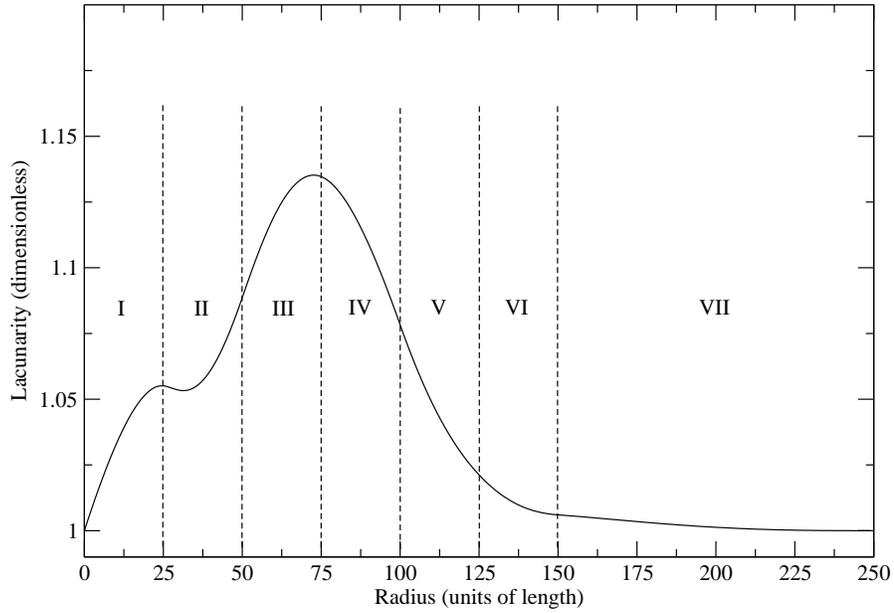} 
\caption{The analytically calculated self-referred lacunarity curve for a
  cross with dimensions $L1$ and $L2$ of $100$ and $50$ units of length respectively.}  
\label{inv_esc_cruz}
\end{figure}

\begin{eqnarray}
A_{I}(x,y,r)  &=& 4L1L2+L2^2 \quad x\in[0,L2],\quad y\in[-L2,0] \nonumber \\
A_{II}(x,y,r) &=& 4L1L2+L2^2 \quad x\in[0,L2],\quad y\in[0,r-L1-L2] \nonumber \\
A_{III}(x,y,r)&=& 3L1L2+L2^2+(r-y-L2)L2 \quad x\in[0,L2],\quad y\in[r-L1-L2,L1] \nonumber 
\end{eqnarray}

\[\mu_{VII}(r)={\frac {31}{9}}\,{L}^{2}+{\frac {20}{9}}\,rL-\frac{2}{9}\,{r}^{2}\]

\[\sigma^2_{VII}(r)=-\frac{4\left( 2\,L-r \right) \left( 5\,L-r \right)^{3}}{81}\]

\[\Lambda_{VII}(r)=-{\frac {3L \left( 13\,{L}^{3}-780\,r{L}^{2}+4\,{r}^{3}+48\,{r}^{2}L
    \right) }{ \left( 31\,{L}^{2}+20\,rL-2\,{r}^{2} \right) ^{2}}}\]
Sumarizing the results for the cross, the whole expression for the lacunarity of a cross is
given by the following expression
\begin{eqnarray} 
 \Lambda(r) = \left\{ 
 \begin{array}{ll}
 \Lambda_I(r)=\frac{4L2(-300Lr+59r2+324L2)}{(36L^{2}-20Lr+3r2)2} 
 \\ \\

 \Lambda_{II}(r)={\frac {{L}^{2} \left( -12\,{L}^{5}r-192\,{r}^{5}L+452\,{L}^{3}{r}^{3}
+{L}^{6}-36\,{L}^{4}{r}^{2}+12\,{r}^{6}+168\,{r}^{4}{L}^{2} \right) }{
{r}^{4} \left( -20\,rL+36\,{L}^{2}+3\,{r}^{2} \right) ^{2}}} 
\\ \\

\Lambda_{III}(r)=-3\,{\frac {64\,{r}^{4}+320\,{r}^{2}{L}^{2}-348\,L{r}^{3}-204\,{L}^{3}
r+37\,{L}^{4}}{ \left( 12\,rL+{L}^{2}+6\,{r}^{2} \right) ^{2}}} 
\\ \\

\Lambda_{IV}(r)=3\,{\frac {L \left( 256\,{r}^{2}L+28\,{r}^{3}-228\,r{L}^{2}+71\,{L}^{3
} \right) }{ \left( 12\,rL+{L}^{2}+6\,{r}^{2} \right) ^{2}}} 
\\ \\

 \Lambda_{V}(r)=\frac {{L}^{2} \left( 1605\,{L}^{6}+5412\,{L}^{4}{r}^{2}+1662\,{r}^{4
}{L}^{2}+40\,{r}^{6}-432\,{r}^{5}L-3320\,{L}^{3}{r}^{3}-4716\,{L}^{5}r
 \right) }{ \left( -100\,{L}^{3}r+90\,{r}^{2}{L}^{2}-24\,L{r}^{3}+2\,{
r}^{4}+49\,{L}^{4} \right) ^{2}} 
\\ \\

 \Lambda_{VI}(r)=-{\frac {{L}^{2} \left( 4788\,{L}^{4}{r}^{2}+666\,{r}^{4}{L}^{2}-8\,{r
}^{6}-3180\,{L}^{3}{r}^{3}-4284\,{L}^{5}r+2145\,{L}^{6} \right) }{
 \left( -100\,{L}^{3}r+90\,{r}^{2}{L}^{2}-24\,L{r}^{3}+2\,{r}^{4}+49\,
{L}^{4} \right) ^{2}}} 
\\ \\

\Lambda_{VII}(r)=-{\frac {3L \left( 13\,{L}^{3}-780\,r{L}^{2}+4\,{r}^{3}+48\,{r}^{2}L
 \right) }{ \left( 31\,{L}^{2}+20\,rL-2\,{r}^{2} \right) ^{2}}} 
\end{array} \right. \nonumber
\end{eqnarray}
The corresponding intervals, I-VII, are shown in Tabel~\ref{tab:regions} and
the resulting lacunarity curves for this piecewisely determined function is
plotted in Figure~\ref{inv_esc_cruz}.

\begin{table}[!t]~\label{tab:regions}
\begin{center}
\begin{tabular}{|l|l|} \hline
Region & Interval\\\hline
I   & $r \in [0,\frac{L2}{2}]$   \\\hline 
II  & $r \in [\frac{L2}{2},L2]$  \\ \hline
III & $r \in [L2,\frac{L1+L2}{2}]$  \\ \hline
IV  & $r \in [\frac{L1+L2}{2},L1]$  \\ \hline
V   & $r \in [L1,L1+\frac{L2}{2}]$  \\ \hline
VI  & $r \in [L1+\frac{L2}{2},L1+L2]$  \\ \hline
VII & $r \in [L1+L2,2L1+L2]$ \\\hline
\end{tabular}
\end{center}
\caption{Regions for the analytical calculation of the self-referred lacunarity.}
\end{table}

\end{document}